\documentclass[prl,aps,preprint,floatfix,showpacs]{revtex4}
\usepackage{graphicx}

\begin{document}

\title{Experimental observation of bias-dependent non-local Andreev reflection}
\author{S. Russo}
\author{M. Kroug}
\author{T. M. Klapwijk}
\author{A. F. Morpurgo}
\address{Kavli Institute of Nanoscience, Delft University of
Technology, Lorentzweg 1, 2628 CJ Delft, The Netherlands}

\date{\today}

\begin{abstract}
We investigate transport through hybrid structures consisting of
two normal metal leads connected via tunnel barriers to one common
superconducting electrode. We find clear evidence for the
occurrence of non-local Andreev reflection and elastic cotunneling
through superconductor when the separation of the tunnel barrier
is comparable to the superconducting coherence length. The
probability of the two processes is energy dependent, with elastic
cotunneling dominating at low energy and non-local Andreev
reflection at higher energies. The energy scale of the crossover
is found to be the Thouless energy of the superconductor, which
indicates the phase coherence of the processes. Our results are
relevant for the realization of recently proposed entangler
devices.
\end{abstract}

\maketitle

\newpage

\begin{figure}[h]
    \centering
    \includegraphics[width=0.5\columnwidth]{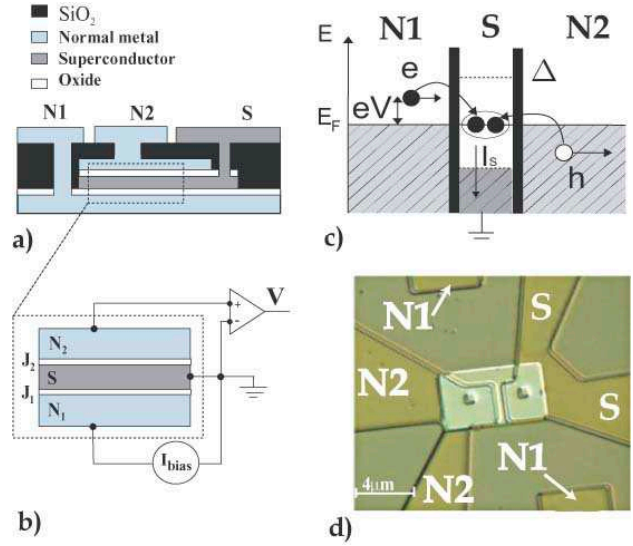}
    \caption  {a) Schematic cross-section of our sample (not to scale).
Two normal electrodes (N1 and N2) are connected to a
superconducting layer (S) via two tunnel barriers (J1 and J2),
whose separation $d$ is defined by the thickness of the
superconducting layer. The concept of the measurement
configuration is shown in b): current is injected through J1 and
the non-local voltage is measured across J2. c) illustrates the
non-local AR process: an incoming electron from N1 is transmitted
as a hole into N2 while a Cooper pair condenses in S. d) Optical
microscope image of one of our samples (top view). The rectangle
in the center is where J1 and J2 are located; N1, N2 and S label
the electrical contacts to the respective metallic layers.}
 \label{Russo_fig1}
 \end{figure}

\begin{figure}[h]
    \centering
    \includegraphics[width=0.5\columnwidth]{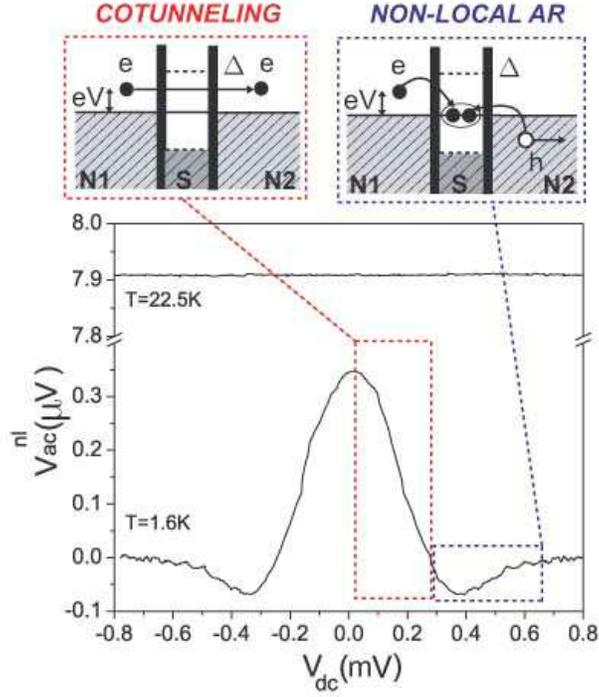}
    \caption  {The non-local voltage $V_{ac}^{nl}$
measured across J2, on a device with a $d=15$ nm thickness of the
superconducting layer, for two different temperatures. The upper
curve is measured at $T = 22.5$ K -well above $T_c^{S}$- and shows
a bias-independent non-local voltage due to electrons. At $1.6 K$
(below $T_c^{S}$), the non-local voltage is much smaller and
depends on the bias $V_{dc}$ across J1. At low bias, $V_{ac}^{nl}$
has the same sign measured in the normal state, indicating that
elastic cotunnelling dominates. At higher bias, the sign of
$V_{ac}^{nl}$ is reversed, which indicates the occurrence of
non-local AR. } \label{Russo_fig2}
\end{figure}

\begin{figure}[h]
    \centering
    \includegraphics[width=0.5\columnwidth]{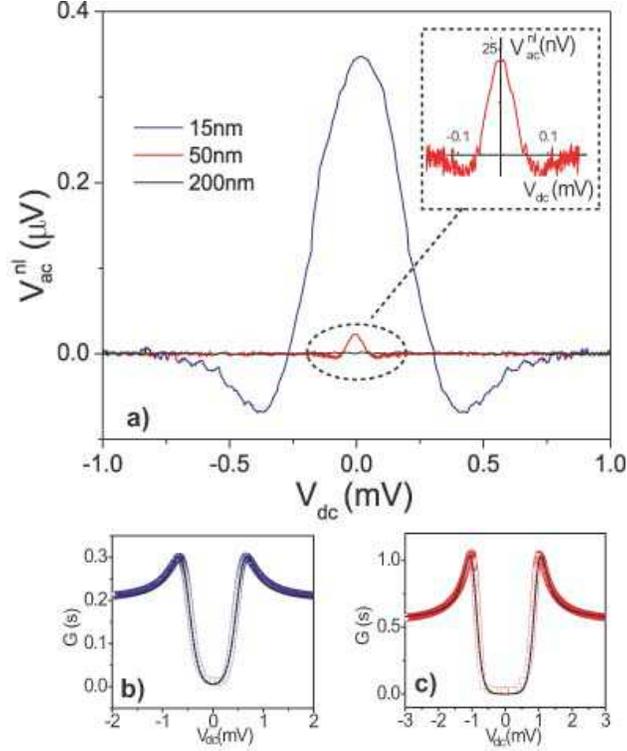}
    \caption  {(a) Non-local voltage $V_{ac}^{nl}$ measured at $T=1.6$ K
on three samples with different thickness of the superconducting
layer ($d=15, \ 50, \ 200$ nm, with a normal state resistance of
$4.8$, $1.7$, and $0.9 \ \Omega$ respectively). Panels b) and c)
show the tunnelling characteristics of junctions, measured in two
devices with $d=15$ and $50$ nm respectively. The solid line is a
fit based on the BCS density of states and shows that good
agreement is found with $\Delta = 0.9$ and $1.45$ mV for the two
different thicknesses of the Nb layer\cite{Long}. The suppression
of the gap in the $d=15$ nm sample is typical of these thin
superconducting films\cite{Floet}.} \label{Russo_fig3}
\end{figure}

Andreev reflection (AR) is a well-known process that enables
charge transfer across an interface between a normal metal and a
superconductor\cite{Andreev}. At this interface, an incoming
electron in the normal metal pairs with a second electron to enter
the superconductor, resulting in a reflected hole. Past work has
focused on the case of holes that are reflected back into the same
electrode from which the incoming electrons originate. However,
recent theoretical studies have considered the possibility that
holes are reflected into a second, spatially separated electrode
\cite{Falci,Bena,Chtchelkatchev, recher3,recher4,feinberg6,
Pistolesi,Prada,Samuelsson,Byers}. It was shown that this \lq\lq
non-local AR" process is equivalent to injecting two
spin-entangled electrons forming the singlet state of a Cooper
pair into two different normal leads\cite{nota2}. In this way,
non-local AR enables the realization of solid-state
entanglers\cite{Burkard} -electronic devices capable of sourcing
entangled pairs of electrons into nano-electronic circuits- that
are of interest for quantum information
processing.\\
One way to investigate the occurrence of non-local AR relies on
the following idea. Two normal metal electrodes are connected via
two tunnel barriers (junctions J1 and J2) to one common
superconducting electrode. If the separation of J1 and J2 is
comparable to the superconducting coherence length $\xi$, an
electron injected at energy $E < \Delta$ from the normal electrode
of J1 can propagate as an evanescent wave through the
superconductor and pair with an electron in the normal electrode
of J2\cite{feinberg6}. This process results in a hole \lq\lq
reflected" into the second electrode, i.e. non-local AR. As holes
have the opposite charge of electrons, holes undergoing non-local
AR generate a voltage difference across J2 that has a sign
opposite to that observed when the superconductor is in the normal
state ($T > T_c^{S}$). Therefore, in principle, the detection of
non-local AR is straightforward: J1 is used to inject current into
the superconductor and J2 is used as a voltage probe to detect a
voltage of the correct sign.\\
In practice, the situation is complicated by the occurrence of a
second process competing with non-local AR: electrons injected
from J1 can be transmitted into J2 without being converted into
holes. This process is known as elastic cotunneling
(EC)\cite{Falci,Pistolesi} and contributes to generate a voltage
across J2 that has the same sign as that observed when the
superconductor is in the normal state. Thus, the sign of the
voltage measured across J2 depends on whether EC or non-local AR
occurs with larger probability. The voltage measured at J2 may
also vanish, if cotunneling and non-local AR occur with exactly
the same probability for all energies of the injected electrons.
As some recent calculations predict\cite{Falci,Pistolesi} that
this could in fact happen, it is not possible to anticipate which
signal -if any- will be measured experimentally. For instance, in
a recent experiment in which two ferromagnetic leads were used as
normal electrodes, only the sign corresponding to EC has been
observed\cite{Beckmann}.\\
In this paper we report a clear experimental evidence for both
non-local AR and elastic cotunneling using the experimental
strategy just outlined. We find that the magnitude and the sign of
the measured non-local voltage depend on the bias across the
injecting junction. At low bias, the observed sign is the same as
when the superconductor is in the normal state, indicating that EC
dominates. At higher bias the sign of the voltage is reversed,
which indicates the occurrence of non-local AR. The energy scale
on which the sign-reversal takes place corresponds to the Thouless
energy of the superconducting layer. From this we conclude that
the sub-gap microscopic processes of conduction, non-local AR and
EC,
are phase-coherent.\\
A schematic representation of the devices used in our experiments
is shown in Fig. \ref{Russo_fig1}a. The structure is implemented
in a Nb/Al multilayer sputtered on a thermally oxidized Si
substrate using conventional Nb/Al technology\cite{Gurvitch}. The
multilayer consists of two normal metal layers (N1 and N2, 50nm Al
layers) connected via two tunnel barriers to one common
superconducor (S). Junction J1 is obtained by {\it in-situ}
oxidation of the N1 layer and subsequent deposition of Nb. Next, a
thin ($\sim \ 5$ nm) Al layer is sputtered on top of the Nb and
oxidized {\it in-situ}. Finally the
top Al layer (N2) is deposited to form junction J2.\\
The fabrication process used to pattern the multilayer relies on
conventional photolithography combined with chlorine-based
reactive ion etching. The junctions area is approximately $4 \
\times \ 8 \ \mu$m$^2$. Independent electrical connections to the
three layers are formed by deposition of a 200 nm thick Al/Nb
layer on a SiO$_{2}$ mask followed by dry etching. We have checked
the quality of the tunnel junctions by fitting the differential
conductance with the usual BCS theory and found that the
tunnelling characteristics of junctions J1 and J2 do not show any
substantial difference. This indicates that the superconducting
properties of the Nb/Al layer (S) are uniform
across its thickness.\\
In our devices the separation between the two tunnel barriers is
determined by the thickness of the S layer, which can be
controlled on the nanometer scale. This is crucial, since the
separation of the tunnel barrier has to be comparable to the
superconducting coherence length in S, $\xi \simeq \sqrt{\xi_0
l_e} \ = \  10 \ - \ 15$ nm \cite{Floet} (where $l_e \ = \
3D/v_{f} \ \simeq \ 2$ nm is the elastic mean free path, the
diffusion constant $D \ = \ 1.6 \ cm^{2}/s$ and $\xi_0 \ = \ \hbar
v_F/\pi \Delta$). An optical microscope image of one of our
devices is shown in
Fig.\ref{Russo_fig1}d.\\
All the measurements were performed at $T \ = \ 1.6$ K or higher,
with the aluminum electrodes N1 and N2 in the normal state
($T_c^{Al} \ \simeq \ 1.2$ K). In the experiment we send current
through one of the junctions (e.g., J1) and measure the non-local
voltage $V^{nl}$ across the other junction (J2), while maintaining
the superconductor at ground. The current bias has a \textit{dc}
component and an \textit{ac} modulation with an amplitude of
$1\mu$A at $19.3$Hz, and a lock-in technique is used to measure
the \textit{ac} component of the non-local signal. This
corresponds to measuring the contribution given to the non-local
voltage by only those electrons which have an energy $E=e V_{dc}$,
where $V_{dc}$ is the \textit{dc} voltage across J1.\\
Fig. \ref{Russo_fig2} shows the $V_{ac}^{nl}$ measured as a
function of $V_{dc}$ at two different temperatures (above and
below $T_c$), on a sample in which the superconducting layer is 15
nm thick (approximately equal to $\xi$). At $T=22.5$ K, when the
Nb is in the normal state, the sample can be simply thought of as
a resistance network: the measured signal is large, because of the
resistance of the thin Nb layer, and bias independent.
Microscopically, the signal is due to electrons injected into the
Nb that have a large probability to diffuse into the lead used as
a voltage probe. At 1.6 K the Nb is superconducting and the Al in
the leads is in the normal state. Now the non-local voltage is
much smaller and it depends on $V_{dc}$. Specifically,
$V_{ac}^{nl}$ reverses its sign at $V_{dc} \ =\ 270 \mu$V and
eventually vanishes at $V_{dc} \simeq \ 700 \ \mu$V, thus on a
bias range much smaller than the superconducting gap (900 $\mu$V,
see Fig. \ref{Russo_fig3}b).\\
To investigate if this signal originates from evanescent waves
propagating below the superconducting gap, we have measured the
non-local voltage in samples with different thickness $d$ of the
superconducting layer. Fig. \ref{Russo_fig3} compares the data
measured in three samples where $d \ =$ 15, 50, and 200 nm,
respectively. For the 50 nm sample, a non-local signal reversing
sign with increasing \textit{dc} bias is still visible at a bias
range much smaller than the superconducting gap. However, the
magnitude of the signal is approximately 20 times smaller than for
the sample with $d \ = \ 15$ nm. For the sample with a 200 nm
thick superconducting layer, no non-local signal is observed.
These observations indicate that $V_{ac}^{nl}$ is very rapidly
suppressed with increasing the thickness of the superconductor, as
expected for evanescent waves. \\
The comparison of different samples additionally shows that the
energy scale on which the non-local signal reverses its sign (and
eventually disappears) becomes smaller for a larger separation of
the tunnel barriers. For the $d \ = \ 15$ nm sample the zero
crossing energy is $\simeq \ 300 \ \mu$eV and for the $d \ = \ 50$
nm it is $\simeq \ 50 \ \mu$eV (see Fig. \ref{Russo_fig3}). These
values correspond well to the Thouless energy $E_T \ = \ \hbar D /
d^2$ of the superconducting layers, equal to $E_T \ \simeq \ 450 \
\mu$V and to $E_T \ \simeq \ 45 \ \mu$V for the $d \ = \ 15$ nm
and the $d \ = \ 50$ respectively. The fact that the Thouless
energy determines the behavior of $V_{ac}^{nl}$, indicates that
the signal originates from quantum-mechanically phase coherent
processes. This is to be expected, since the transit time
$\tau_{tr}$ of electrons injected from J1 and transmitted into J2
-as electrons or holes- is $\tau_{tr} \ \simeq \ d^2/D \ \simeq \
1 \ - \ 10$ ps, much smaller than the inelastic electron-phonon
($\tau_{ph} \ \simeq \ 1$ ns at 1 K in Nb) and electron-electron
($\tau_{ee} \ \simeq \ 0.1$ ns) interaction
times\cite{Ptitsina}.\\
Finding that $E_{T}$ is the relevant energy scale in our
measurements also gives an indication that non-equilibrium
effects\cite{NATO} in the superconductor do not play a relevant
role in determining the behavior of $V_{ac}^{nl}$. In fact, these
effects depend on the quasiparticle injection rate and relaxation
times, whose energy dependence is not strongly influenced by phase
coherent propagation in the superconductor (and thus by $E_{T}$).
Note also that non-equilibrium effect normally become more
relevant at higher bias voltage (when the amount of injected
charge is larger), whereas the amplitude of the signal
$V_{ac}^{nl}$ is maximum at $V_{dc}=0$ V and vanishes for $V_{dc}$
well below the gap. The absence of non-equilibrium is consistent
with the low transparency of our tunnel barriers ($T \approx
10^{-5}$) and with the fact that quasiparticles are injected with
energies well below the superconducting gap. In contrast to
quasiparticles occupying states above $\Delta$, which may have
very long relaxation times, quasiparticles with $E < \Delta$ decay
very rapidly on the scale of $h/\Delta$.\\
Having established the absence of significant non-equilibrium
effects, we conclude that the measured non-local voltage
$V_{ac}^{nl}$ is due to phase coherent elastic cotunneling and
non-local AR. EC is predominant at low bias whereas non-local AR
dominates at higher bias, where the sign of $V_{ac}^{nl}$ is
negative. That the effect is large and present in all samples
(approximately 10 samples with d=15nm and 50nm have been studied)
demonstrates that the sign reverse in the non-local voltage is not
just a sample-specific effect, as has been observed in InAs/Nb
structures\cite{den
Hartog}.\\
The measured temperature and magnetic field dependence of
$V_{ac}^{nl}$ (see Fig. 4) are consistent with this
interpretation. $V_{ac}^{nl}$ increases with lowering $T$
similarly to what one would expect from the convolution of a
thermally smeared Fermi distribution with an energy dependent
transmission probability (and excluding the possibility that the
signal is due to quasiparticle propagating above the gap). The
signal is suppressed by a magnetic field applied parallel to the
superconducting layer at $B \ \simeq \ 0.5$ T, which is much
smaller than the critical field of our S layer (higher than $6$
T\cite{nota4}). Since $\Delta$ is only slightly reduced (few
percent) by such a field, we believe that the main effect of $B$
is the breaking of time reversal symmetry for the electron-hole
wave injected into the superconductor. Note, however, that at 0.5
T the magnetic flux enclosed by typical electron-hole trajectories
in the superconductor ($d$ is smaller than the magnetic
penetration length in Nb for all samples) is only approximately
$0.2 \times \phi_0$. \\
Our observation of a non-local signal shows that the cancellation
of the contribution to $V_{ac}^{nl}$ due to non-local AR and EC
does not occur in the samples investigated here. This cancellation
was theoretically found in models that neglect the effect of
Coulomb interaction\cite{Falci}, whereas calculations made for
different systems in which interactions in the leads play a
relevant role \cite{recher4, Bena, recher3} did all predict the
occurrence of visible effects. Since the effect of Coulomb
interaction on electronic transport is visible in large-area
tunnel junctions of size comparable to ours\cite{pierre}, we
believe that Coulomb interaction may also be relevant here. A
quantitative interpretation of our experimental results will
require the analysis of theoretical models more sophisticated than
those considered until now, which may have to address aspects of
our samples that have not been considered so far (e.g. a gradient
in the phase of the superconducting order parameter, or a small
sub-gap density of states induced by
the presence of the normal electrodes).\\
In conclusion, we have reported clear experimental evidence for
the occurrence of non-local Andreev reflection and elastic
cotunneling through a superconducting layer. Our results show hat
these processes are phase coherent and strongly depend on the
energy of the injected electrons. These findings are relevant for
recent theoretical proposals of quantum entangler devices that aim
at injecting into two spatially separated normal metal leads the
spin-entangled electrons forming a Cooper pair. In this context,
the energy dependence of the probability for non-local Andreev
reflection may provide a
new way to control the output of these entanglers.\\
The authors acknowledge helpful discussions with Y. V. Nazarov and
P. Samuelsson. This work was financially supported by NWO/FOM and
by NOVA. The work of AFM is part of the NWO Vernieuwingsimpuls
2000 program.


\begin{thebibliography}{99}
\bibitem{nota2} This is a central concept of all the theoretical work referred to above.
For particularly detailed discussions see, e.g., Refs. 3,8, and
10.
\bibitem{nota4} The presence of superconductivity for B up to
6 T is visible in the tunneling characteristics of J1 and J2.
\end{thebibliography}

\newpage

\newpage

\newpage

\newpage

\begin{figure}[h]
    \centering
    \includegraphics[width=0.5\columnwidth]{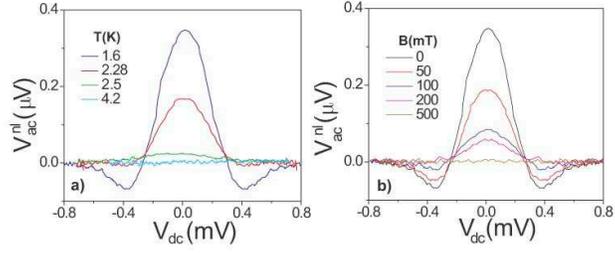}
    \caption  {(a) Temperature and (b) magnetic field dependence of the
non-local voltage $V_{ac}^{nl}$ measured as a function of
$V_{dc}$, on a sample with $d=15$ nm.}
  \label{Russo_fig4}
\end{figure}

\end{document}